\documentclass[12pt,pra]{revtex4}
\usepackage{times}
\usepackage{t1enc}
\usepackage[english]{babel}
\usepackage{graphicx}

\begin{document}

\title{Slow light transmission in one-dimensional periodic structures}

\author{O. del Barco and M. Ortu\~no}

\affiliation{Departamento de F\'{i}sica - CIOyN, Universidad de
Murcia, Spain}

\begin{abstract}
We have analyzed the transmission properties of pulses through
one-dimensional periodic structures in order to systematically
explore the best conditions to achieve the maximum delay with the
minimum possible distortion. In the absence of absorption and no
layer variation, the transmission coefficient $t_N$ can be well
approximated by a sum of Lorentzian resonances. The ratio between
their width $\Gamma_{\rm r}$ and their separation
$\Delta\omega_{\rm r}$ is a crucial parameter to characterize the
distortion of the transmitted pulse. For typical values of the
parameters used in telecommunications and high index of refraction
contrasts $n_2 / n_1$, the distortion of the transmitted pulse is
unacceptably large for frequencies near the edge of the
transmission window. We estimate fractional delays achievables in
terms of the central frequency used and the pulse bandwidth.
\end{abstract}

\maketitle

\section{Introduction}

Slow light phenomena has recently attracted a great deal of
interest. Many researchers have delayed optical pulses in a great
variety of physical systems such as atomic gases
\cite{HAU99,LIU01,PHI01}, optical fibers
\cite{OKAW05,SONG05,POV05} or photonic-crystal devices
\cite{NOT01,GER05,MOR04,VLA05}.

The ability of delaying optical pulses has many potential
applications in the field of high-speed optical processing such as
random-access memory, data synchronization or pattern correlation.
In these communications applications, where large bandwidth
systems are required in order to respond fast to the short pulses
which transport the information, the pulse of light must be
delayed several times the pulse duration without significant
distortion.

Recently, Mok \emph{et al.} \cite{MOK06} managed to delay optical
pulses through the use of solitons launched near the band gap edge
of a fiber Bragg grating (FBG). They demonstrated the propagation
of 0.68 ns optical pulses travelling at $v_{\rm g} = 0.23/c$ in a
10 cm FBG leading to delays corresponding to about 2.4 pulse
widths. Sarkar \emph{et al.} \cite{SAR06} obtained a fractional
pulse delay of 200 \% for 8 ps pulses in a $\rm{GaAs}/
\rm{AlGaAs}$ multiple quantum well sample and Novikova \emph{et
al.} \cite{NOV08} managed to slow light in Rb vapour with minimal
loss and fractional pulse delays greater than 10. In a numerical
simulation of light propagation, Boyd \emph{et al.} \cite{BOYD05}
obtained a delay of 75 pulse lengths under realistic laboratory
conditions.

Our group has studied the traversal time of electronic and
photonic pulses concentrating on the case of exponentially
decaying wave functions, i.e., tunneling experiments in electronic
and frequency gaps in optical transmission \cite{BA06}. Our
previous aim was to study how fast a wave could travel in a given
region. For that purpose, we developed a method to calculate the
traversal time including finite size effects, i.e., taking into
account the specific form of the pulse. We found that the
traversal time $\tau$ for an incident wave packet with Fourier
components $\Phi_{\rm i}(\omega)$ which traverses a structure of
length $L$ can be written as \cite{BA06}
\begin{equation}\label{exptempus}
\tau = \frac{\int_{0}^{\infty} d\omega \ |t(\omega)|^2 \
|\Phi_{\rm i}(\omega)|^2 \ \tau_1(\omega)}{\int_{0}^{\infty}
d\omega \ |t(\omega)|^2 \ |\Phi_{\rm i}(\omega)|^2},
\end{equation}
where $|t(\omega)|^2$ is the transmission coefficient and
$\tau_1(\omega)$ is the so called B\"uttiker--Landauer time or
phase time, which corresponds to the frequency derivative of the
phase of the complex transmission amplitude $\varphi_{\rm
t}(\omega)$ \cite{GO95}
\begin{equation}\label{ty}
\tau_1(\omega) = {\partial\varphi_{\rm t}(\omega)\over\partial
\omega}.
\end{equation}

In this paper we analyze under which conditions we can slow down
light pulses as much as possible (large fractional delays) with
minimum distortion in one-dimensional (1D) periodic structures,
such as FBGs, usually employed in optical communication systems.
The most important parameter to quantify how slow light is
traversing a region is the fractional delay FD of our transmitted
pulse, which is defined as
\begin{equation}\label{fracdelaytempus}
\mbox{FD} = \frac{1}{\Delta t} \left( \tau - \frac{L}{c} \right),
\end{equation}
where $\Delta t$ is the width of the incident pulse. We have used
Eq.\ (\ref{fracdelaytempus}) to calculate the fractional delay of
the transmitted pulse, and we have found that under most
circumstances the classical calculation constitutes an excellent
approximation. Within this approach the delay time is given by
\begin{equation}\label{fracdelayclass}
\mbox{FD}_{\rm cl} = \frac{L}{\Delta t} \left( \frac{1}{v_{\rm g}} -
\frac{1}{c} \right),
\end{equation}
where $v_{\rm g}$ is the group velocity of the transmitted pulse.

As the group velocity tends to zero as we approach the edge of the
transmission window produced by a periodic structure, it is
natural to use frequencies near this edge to slow down light as
much as possible. However, we have found that, for high index of
refraction contrasts $n_2 / n_1$, the pulse distortion is
unacceptably large near the edge and usually does not compensate
to work in this region. For a given degree of distortion and high
contrasts, it is better to use short pulses near the center of the
frequency window to obtain large values of the FD.

We employ the characteristic determinant method \cite{AG91} to
calculate the transmission coefficient through a layered periodic
structure \cite{GO95a}. This method constitutes an excellent tool
to derive analytical expressions and to obtain numerical results
for all the relevant quantities in periodic arrangements. We have
considered an idealized case with no absorption and no layer
variation. The inclusion of any of these effects is difficult and
outside the scope of this work, although it will be the subject of
future work. In the discussions, we set the limits of validity of
our present model and discuss possible effects due to absorption
and layer variability.

The plan of the work is as follows. In Sec.\ II we present the 1D
periodic structure used in our numerical calculations and describe
in detail its transmission coefficient obtained with the
characteristic determinant method. In Sec.\ III we study the pulse
propagation through the 1D grating depending on the position of
its central frequency in the transmission window, and in Sec.\ IV
we calculate numerically the transmission coefficient and the
fractional delay of our pulse as a function of the frequency
bandwidth. We also analyze the convenience of using central pulse
frequencies near the center or the edge of the transmission
window. Finally, we summarize our results in Sec.\ V and present
an Appendix with details of the calculation of the transmitted
pulse through a periodic arrangement of Lorentzian resonances.

\section{Transmission coefficient analysis of the 1D periodic structure}

\begin{figure}
\includegraphics[width=.5\textwidth]{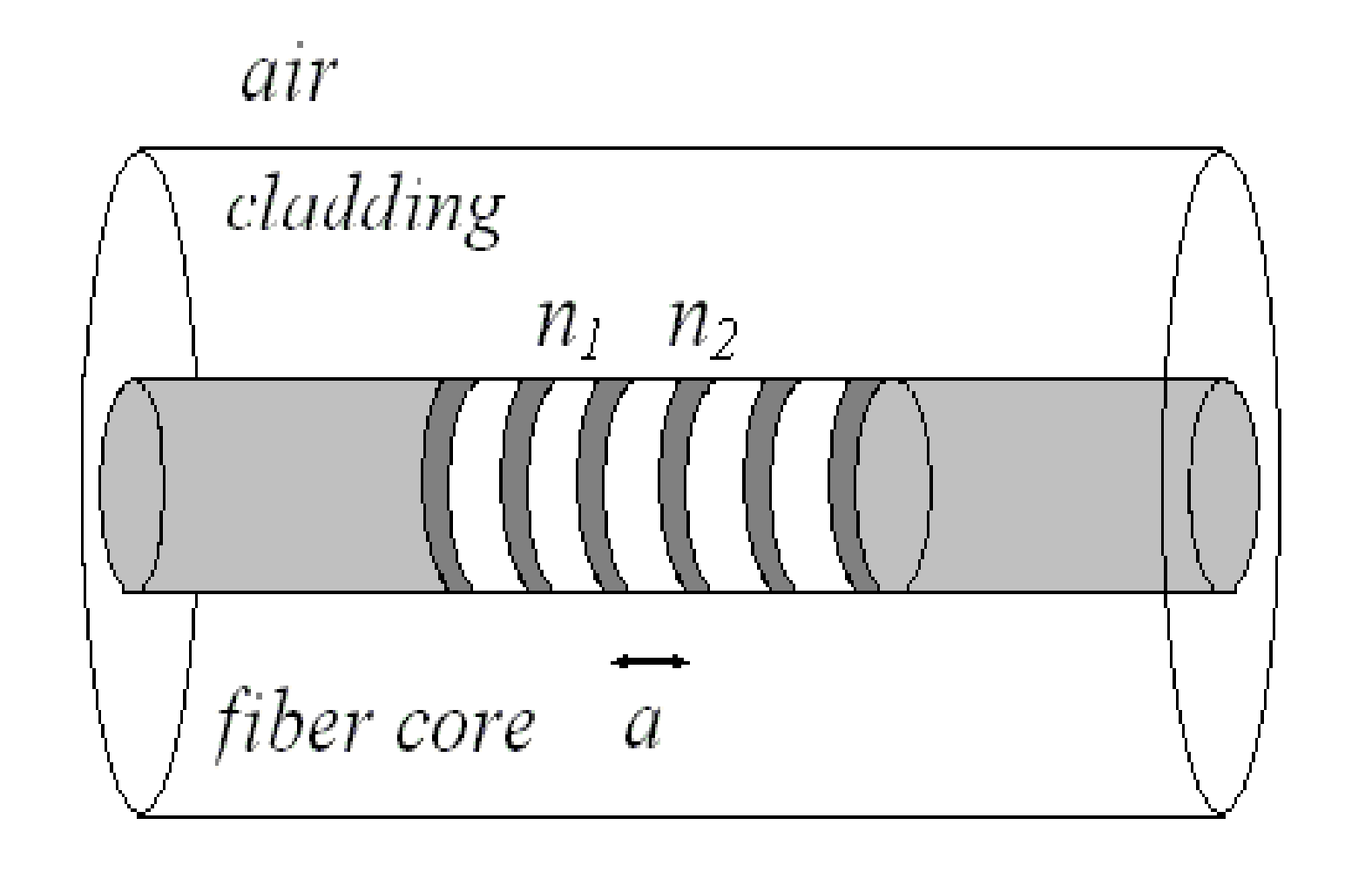}
\caption{A fiber Bragg grating with an uniform periodic modulation
of the index of refraction.}\label{fig1}
\end{figure}

The slow light structure considered throughout the paper is a
periodic arrangement of layers with index of refraction $n_1$ and
thickness $L_1$ alternating with layers of index of refraction
$n_2$ and thickness $L_2$. The wavenumbers in layers of both types
are $k_i = \omega n_i /c$ for $i=1, 2$ and the grating period is
$a = L_1 + L_2$. We have chosen $n_1L_1=n_2L_2$. The periodic
modulation of the index of refraction can be constructed in the
core of a short segment of optical fiber constituting a fiber
Bragg grating. Such structures present a photonic band gap, where
light pulses are totally reflected, and are typically used as
inline optical filters (see Fig.\ \ref{fig1}).
In our numerical calculations the FBG consists of 40000 Si layers
of length 158.8 nm and index of refraction 3.48 alternating with
39999 air layers of 553.2 nm wide. The grating period is then
712.0 nm and the total size of our arrangement 28.5 mm. This
corresponds to a realistic structure used in slow light
experiments \cite{POV05,MOK06}, but the results are of general
validity.

To obtain the transmission coefficient of the previous structure
$|t_N|^{2}$, we employ the characteristic determinant method
\cite{AG91}. This formalism allows one to express the transmission
coefficient of a wave propagating in a 1D periodic structure
through the determinant $D_N$, which depends on the amplitudes of
transmission and reflection of a single cell. Via this determinant
method, the inverse of the transmission coefficient $|t_N|^{-2}$ for this
structure can be written as \cite{CU96}
\begin{equation}\label{invtranscoef}
|t_N|^{-2} = 1 + \left[ {\frac{k_1^2-k_2^2}{2k_1k_2}} \sin (k_2
L_2)\right]^2 \frac{\sin^2(N\beta a /2)}{\sin^2(\beta a)},
\end{equation}
where $N-1$ is the total number of layers and $\beta$ plays the
role of quasimomentum of the system, and is defined by
\begin{equation}\label{beta}
\cos (\beta a) = \cos (k_1 L_1)\cos (k_2 L_2) - \left(
{\frac{k_1^2+k_2^2}{2k_1k_2}} \right) \sin(k_1 L_1) \sin(k_2 L_2).
\end{equation}
When the modulus of the RHS of Eq.\ (\ref{beta}) is greater than
1, $\beta$ has to be taken as imaginary. This situation
corresponds to a forbidden frequency window.

\begin{figure}
\includegraphics[width=.6\textwidth]{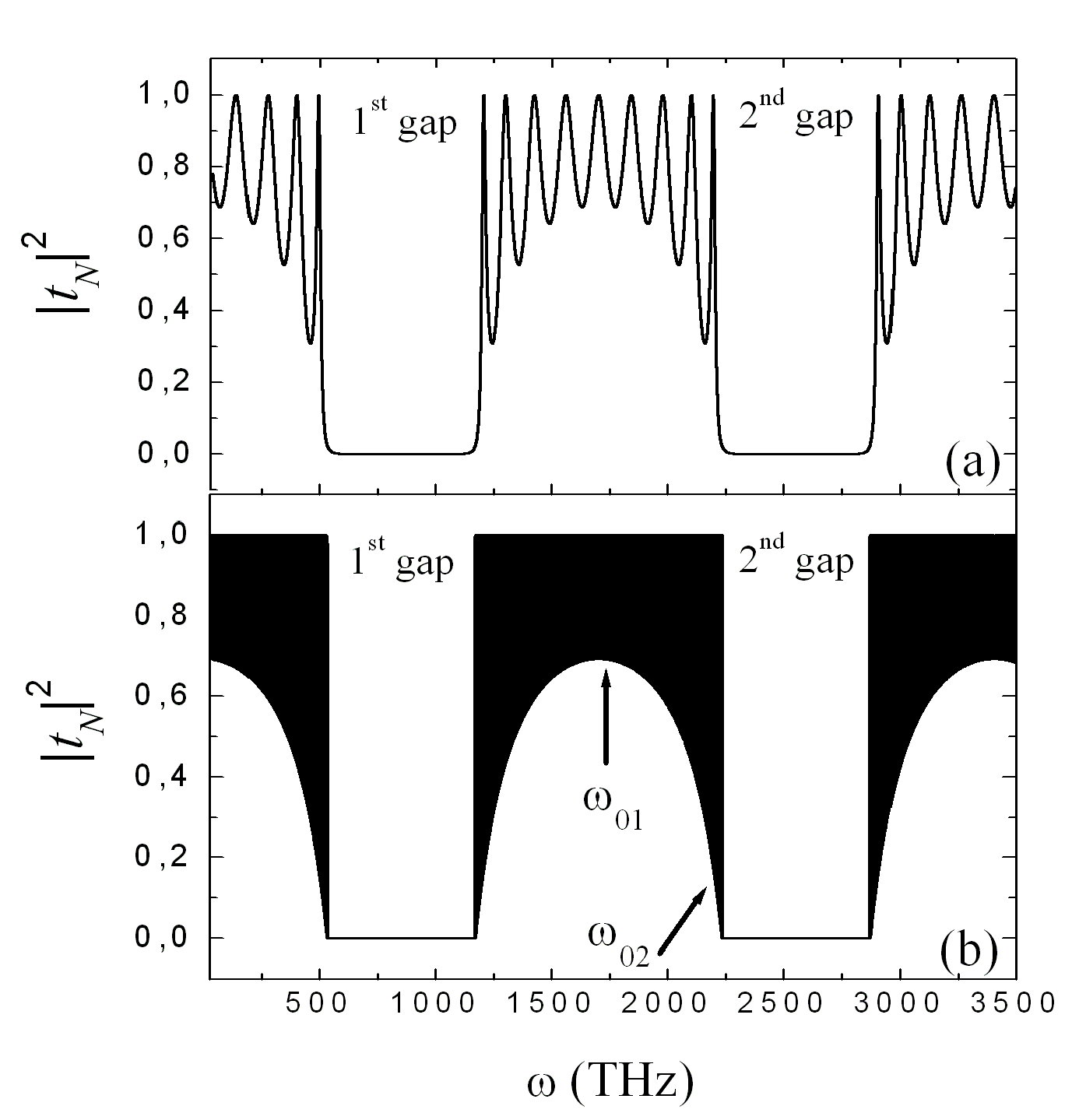}
\caption{Transmission coefficient $|t_N|^2$ as a function of the
frequency $\omega$ for (a) $N=10$ and (b) $N=80000$. The regions
around the frequencies $\omega_{01} = 1754$ THz and $\omega_{02} =
2225$ THz marked with arrows are expanded in Fig.\
\ref{fig3}.}\label{fig2}
\end{figure}

\begin{figure}
\includegraphics[width=.75\textwidth]{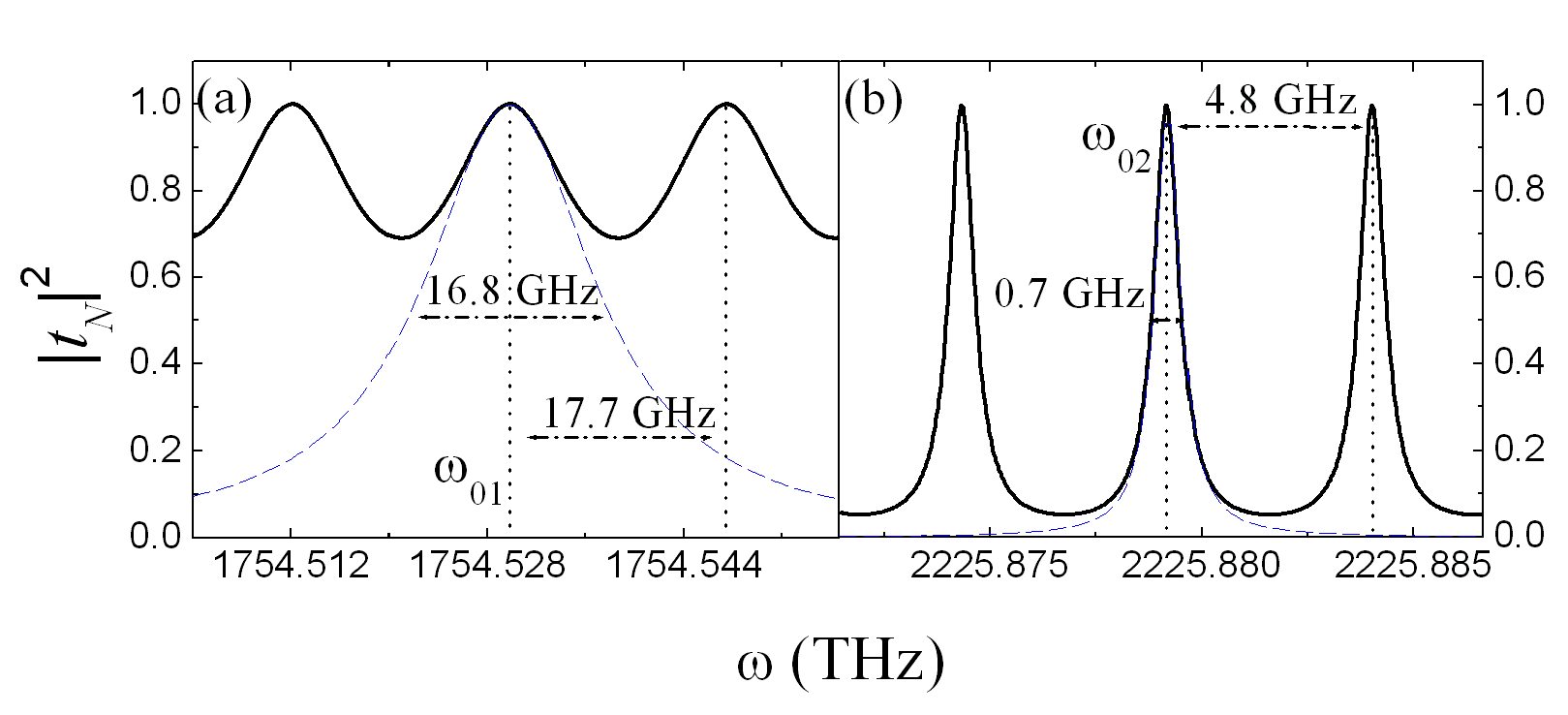}
\caption{Transmission coefficient $|t_N|^2$ versus
$\omega$ for a region near the center (a) and the edge (b) of the
transmission window plotted in Fig.\ \ref{fig2}(b). The dashed
curves are Lorentzian fits for a single peak.}\label{fig3}
\end{figure}

In Fig.\ \ref{fig2}(a) we have represented the transmission
coefficient of our grating $|t_N|^2$ as a function of the
frequency $\omega$ for $N=10$ to illustrate its behavior. The
realistic grating corresponds to Fig.\ \ref{fig2}(b) where
$N=80000$ (with 28.5 mm length). There are so many peaks that they
appear as a black region in the figure, and they cannot be
resolved in the scale shown.
In order to study $|t_N|^2$ in detail as a function of the
position in the frequency transmission window we plot on a much
larger scale a region near the center, Fig.\ \ref{fig3}(a), and a
region near the edge, Fig.\ \ref{fig3}(b). The center of these two
regions, $\omega_{01} = 1754$ THz and $\omega_{02} = 2225$ THz,
are marked with arrows in Fig.\ \ref{fig2}(b) and they correspond
to wavelengths within the infrared spectrum ($\lambda_{01} =
1073.6$ nm and $\lambda_{02} = 846.3$ nm) for the second
transmission window. In percentage terms, the frequencies
$\omega_{01}$ and $\omega_{02}$ are at 45 \% and 1 \% of the total
window width from the upper frequency window edge, respectively.
Similar results will be obtained for the other frequency windows.

The dashed lines in Fig.\ \ref{fig3}(a) and \ref{fig3}(b)
correspond to Lorentzian fits of the central resonance for each
case. We note that the Lorentzian curves fit fairly well the peaks
in the transmission coefficients in all cases studied.
We can in fact justify this by
having a close look into Eq.\ (\ref{invtranscoef}) where the term
$|t_N|^{-2} - 1$ is the product of a highly oscillating function,
$\sin^2(N\beta a/2)$, and factors independent of the number of
layers and which vary very slowly on the scale of one peak. The
oscillating function can be approximated around each of its
minimum by a quadratic term, resulting in a Lorentzian curve for
the transmission coefficient. We will use this expansion later on
to calculate the width of each peak.

One observes that the resonances are narrower and closer to each
other near the transmission window edge. From now on, we will
denote by $\Gamma_{\rm r}$ the energy width of each individual
Lorentzian resonance and by $\Delta\omega_{\rm r}$ the frequency
separation between two consecutive resonances. The ratio
$\Gamma_{\rm r} / \hbar \Delta\omega_{\rm r}$, which we will see
that constitutes an important parameter of the problem, decreases
as we approach the window edge. For this reason, the overlap
between Lorentzians and the transmission in the midpoint between
peaks is very small in this region.

We can now obtain analytical expressions for the width
$\Gamma_{\rm r}$ and the separation $\Delta\omega_{\rm r}$. The
term within brackets in Eq.\ (\ref{invtranscoef}) only depends on
the properties of one layer, while the quotient of the sine
functions contains information about the interference between
different layers. The transmission coefficient $|t_N|^{2}$ is
equal to 1 when $\sin{(N\beta a/2)} = 0$ and $\beta$ is different
from 0. This condition occurs for
\begin{equation}\label{rescont}
\beta a = \frac{2 \pi n}{N} \qquad n = 1, \dots, N/2 -1,
\end{equation}
and it corresponds to a resonant frequency. Considering that each
resonance is a Lorentzian function centered about the resonant
frequency $\omega_{\rm r}$ with full width $\Gamma_{\rm r} /
\hbar$, we can write the inverse of the transmission coefficient
for one resonance as
\begin{equation}\label{invtranscoeflor}
|t|^{-2} = 1 + \frac{4 \hbar^2}{\Gamma_{\rm r}^2} (\omega -
\omega_{\rm r})^2.
\end{equation}

Expanding the sine in the numerator of Eq.\ (\ref{invtranscoef})
near a resonance we have
\begin{equation}\label{finiteincr}
\sin^2 \left(\frac{N\beta a}{2}\right) \simeq \frac{N^2
(\delta\beta)^2 a^2}{4} \simeq \frac{N^2 a^2}{4 v_{\rm g}^2}
(\delta\omega)^2,
\end{equation}
so, Eq.\ (\ref{invtranscoef}) can be rewritten as
\begin{equation}\label{invtranscoefapp}
|t_N|^{-2} = 1 +  \left[\frac{(k_1^2-k_2^2) \sin (k_2
L_2)}{2k_1k_2 \sin(\beta a)} \right]^2  \frac{N^2 a^2}{4 v_{\rm
g}^2} (\delta\omega)^2,
\end{equation}
where we have taken into account the definition of the group
velocity $v_{\rm g}$ in terms of the quasimomentum $\beta$
\begin{equation}\label{vG}
v_{\rm g} = \frac{\delta\omega}{\delta\beta}.
\end{equation}

We can derive an expression for the frequency distance between two
consecutive resonances $\Delta\omega_{\rm r}$ by taking finite
differences in Eq.\ (\ref{rescont}) (resonance condition)
\begin{equation}\label{sepres}
\Delta\omega_{\rm r} = v_{\rm g} \Delta\beta = v_{\rm g} \frac{2
\pi}{N a},
\end{equation}
and after comparing Eqs.\ (\ref{invtranscoeflor}) and
(\ref{invtranscoefapp}) we can identify
\begin{equation}\label{gamma1res}
\frac{\Gamma_{\rm r}}{\hbar \Delta\omega_{\rm r}} = \left( \frac{2
|\sin(\beta a)|}{\pi} \right) \left[ {\frac{k_1^2-k_2^2}{2k_1k_2}}
\sin (k_2 L_2)\right]^{-1}.
\end{equation}

\begin{figure}
\includegraphics[width=.6\textwidth]{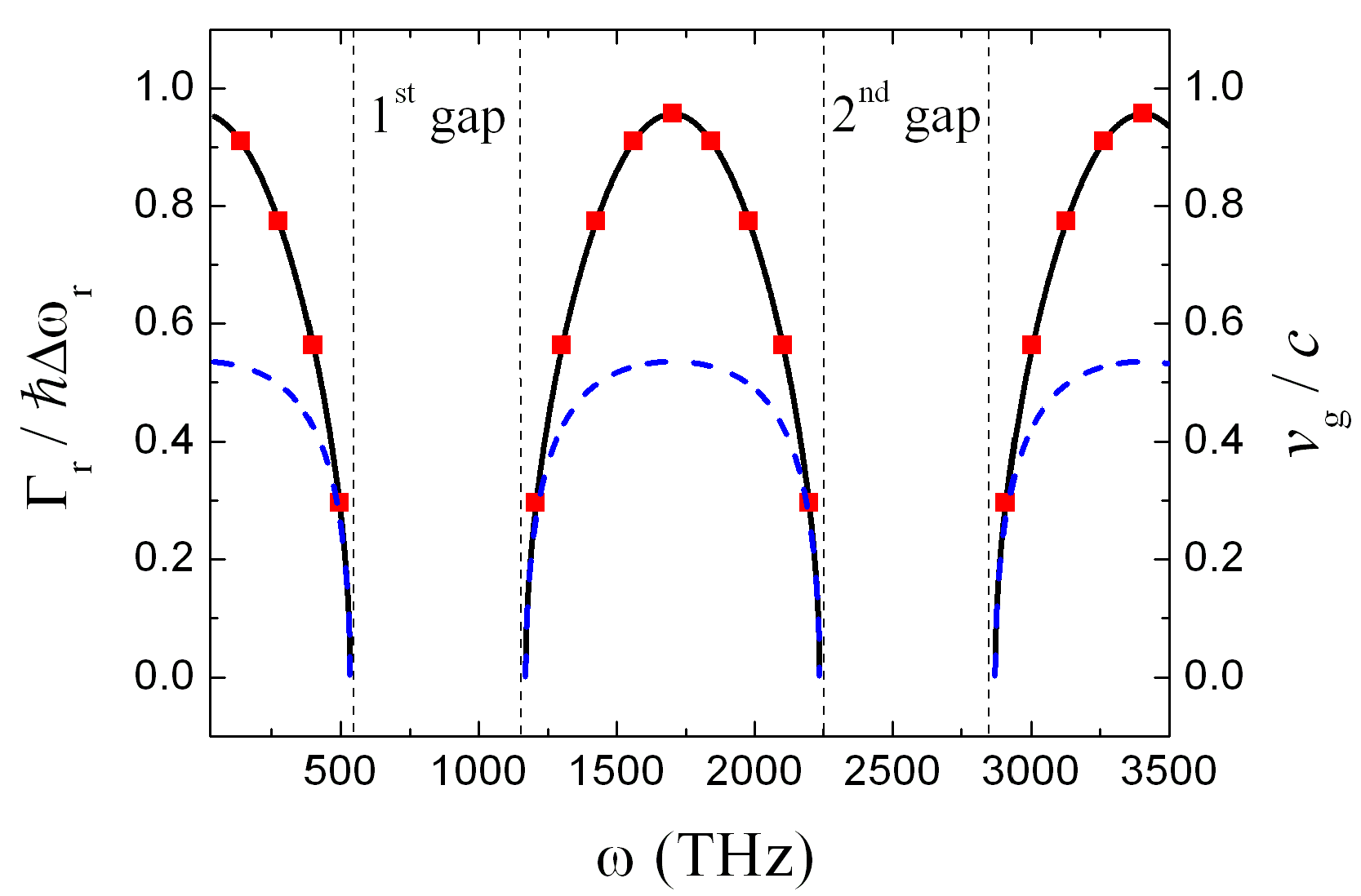}
\caption{The ratio $\Gamma_{\rm r} / \hbar
\Delta\omega_{\rm r}$ (left axis, squares for $N=10$ and solid
line for $N=80000$) and the fractional group velocity $v_{\rm g} /
c$ (right axis, dashed line) versus the frequency
$\omega$.}\label{fig4}
\end{figure}

As we will see, the ratio $\Gamma_{\rm r} / \hbar
\Delta\omega_{\rm r}$ is a very significant parameter which is
independent of the number of layers of our grating. In Fig.\
\ref{fig4} we plot this ratio,  calculated via Eq.\
(\ref{gamma1res}), as a function of the frequency $\omega$ (left
axis, squares for $N=10$ and solid line for $N=80000$). One
observes that in the center of the transmission windows the ratio
$\Gamma_{\rm r} / \hbar \Delta\omega_{\rm r}$ reaches its maximum,
which is roughly 1.0 in our case. This ratio also tends to zero as
we approach the optical mirror region, what has important
consequences on the distortion of the transmitted pulses, as we
will see.

It is well known that the group velocity tends to zero when
approaching the edge of the transmission windows. This is the
reason why most authors have tried to work in this region to slow
down light as much as possible. In Fig.\ \ref{fig4} we also show
the ratio of the group velocity to the speed of light in vacuum
$v_{\rm g} / c$ (right axis, dashed line) versus the frequency
$\omega$. As expected, this ratio tends to zero as we approach the
transmission window edge.

\section{Pulse propagation through the transmission window}

Let us now study how an incident gaussian pulse $\Psi_{\rm
inc}(t)$ traverses our 1D grating depending on the position of its
central frequency in the transmission window. We will consider the
two frequencies described in Fig.\ \ref{fig2}(b), $\omega_{01} =
1754$ THz (near the center) and $\omega_{02} = 2225$ THz (near the
edge). In both cases, we have chosen a pulse width of 68.4 ps
which corresponds to a bandwidth of about 41.3 GHz.

For these broadband pulses we can approximate the transmission
amplitude of our grating by a periodic function of $\omega$ that
can be written as the sum of a given function displaced
periodically by a fixed amount
\begin{equation}\label{transamplappend}
t_N(\omega) = \sum_n f(\omega - \Delta\omega_{\rm r} n,
\Gamma_{\rm r}).
\end{equation}
As before, $\Delta\omega_{\rm r}$ is the frequency separation
between consecutive resonances. As we saw in the previous section,
we can consider, to a very good approximation, $f$ to be a
Lorentzian function of  width $\Gamma_{\rm r}$. Under these
conditions, we have shown in the Appendix that the transmitted
pulse consists of an exponentially decreasing series of similar
pulses given by
\begin{equation}\label{psitr6}
\Psi_{\rm tr}(t) = \frac{\sqrt{2 \pi} C}{\Delta\omega_{\rm r}} \
\exp\left[-\frac{\Gamma_{\rm r} |t|}{2 \hbar}\right] \ \sum_n
\Psi_{\rm inc} \left( t - \frac{2 \pi n}{\Delta \omega_{\rm r}}
\right),
\end{equation}
where $C$ is a normalization constant. If we neglect overlap
between consecutive peaks, it is easy to deduce from the previous
expression that the ratio $r$ between the intensity of two of
these peaks is equal to
\begin{equation}\label{psitr7}
r = \exp\left[-\frac{2\pi\Gamma_{\rm r}}{\hbar \Delta\omega_{\rm
r}}\right].
\end{equation}
According to this result, Eq.\ (\ref{psitr7}), Fig.\ \ref{fig4}
which represents $\Gamma_{\rm r}/\hbar \Delta\omega_{\rm r}$ also
corresponds to the frequency dependence of the factor $r$ on a
logarithmic scale.

\begin{figure}
\includegraphics[width=.5\textwidth]{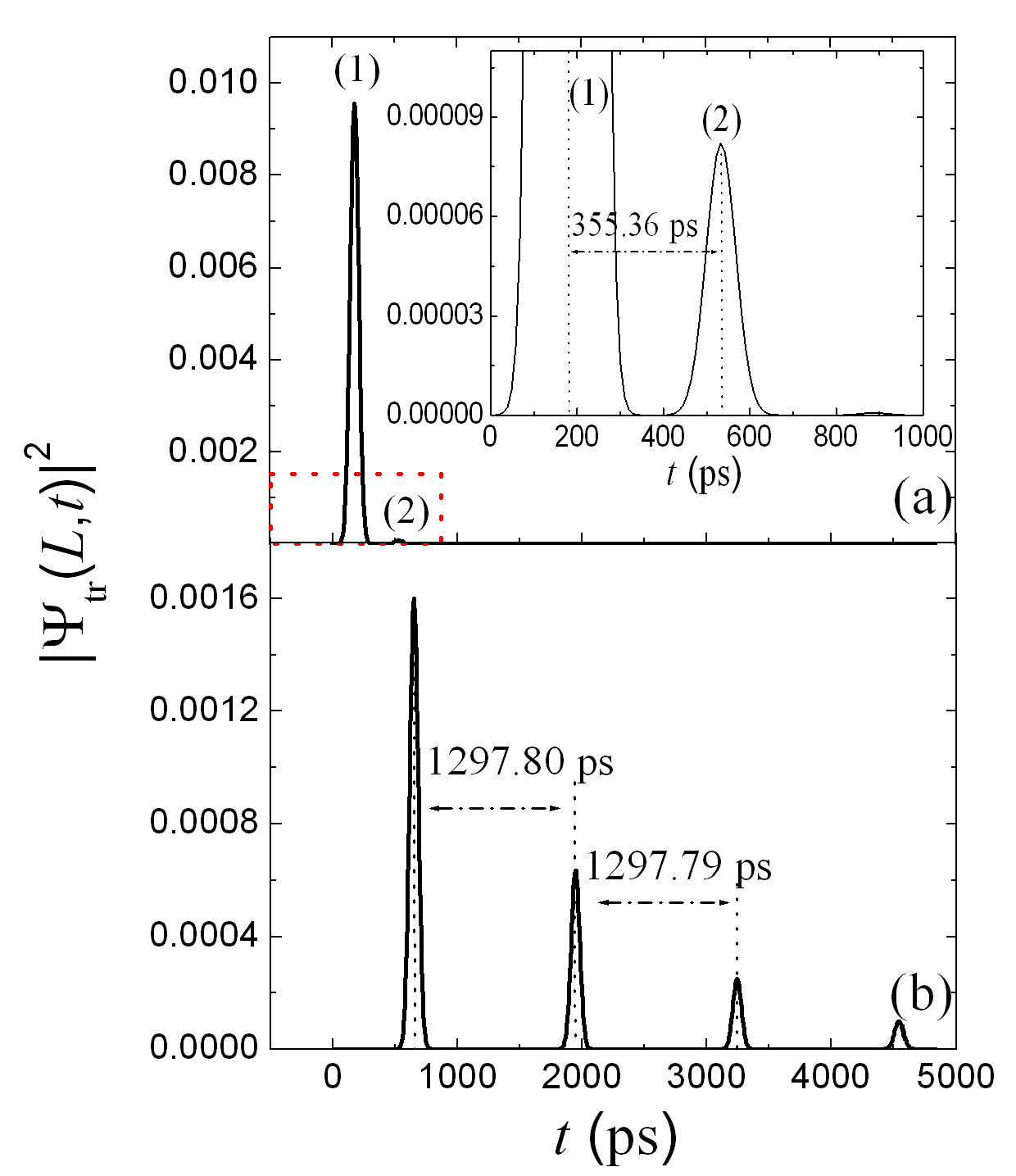}
\caption{Transmitted pulses through our 1D grating
for an incident gaussian pulse of width 68.4 ps and a central
frequency equal to $\omega_{01} = 1754$ THz (a) and $\omega_{02} =
2225$ THz (b). The inset in (a) corresponds to the same curve as
in the main part with a different vertical scale.}\label{fig5}
\end{figure}

Near $\omega_{01}$ the ratio $r$ is very small (about a $0.25 \%$)
and the secondary peak cannot be appreciated in the main part of
Fig.\ \ref{fig5}(a). In the inset we show the same data on a large
vertical scale, in order to appreciate more clearly this secondary
peak, whose temporal separation with the principal peak is about
355.36 ps. Near $\omega_{02}$ the parameter $\Gamma_{\rm r} /
\hbar \Delta\omega_{\rm r}$ is lower than in the previous case and
the ratio $r$ is greater (roughly 40 \%) and several peaks can be
appreciated, as in Fig.\ \ref{fig5}(b). It is clear that near the
transmission window edge the importance of secondary peaks is
significant. Their separation, 1297.80 ps, is larger than in the
center of the window because this is inversely proportional to the
frequency distance between two consecutive Lorentzian resonances
$\Delta\omega_{\rm r}$ (see Eq.\ (\ref{psitr6})).

\begin{figure}
\includegraphics[width=.6\textwidth]{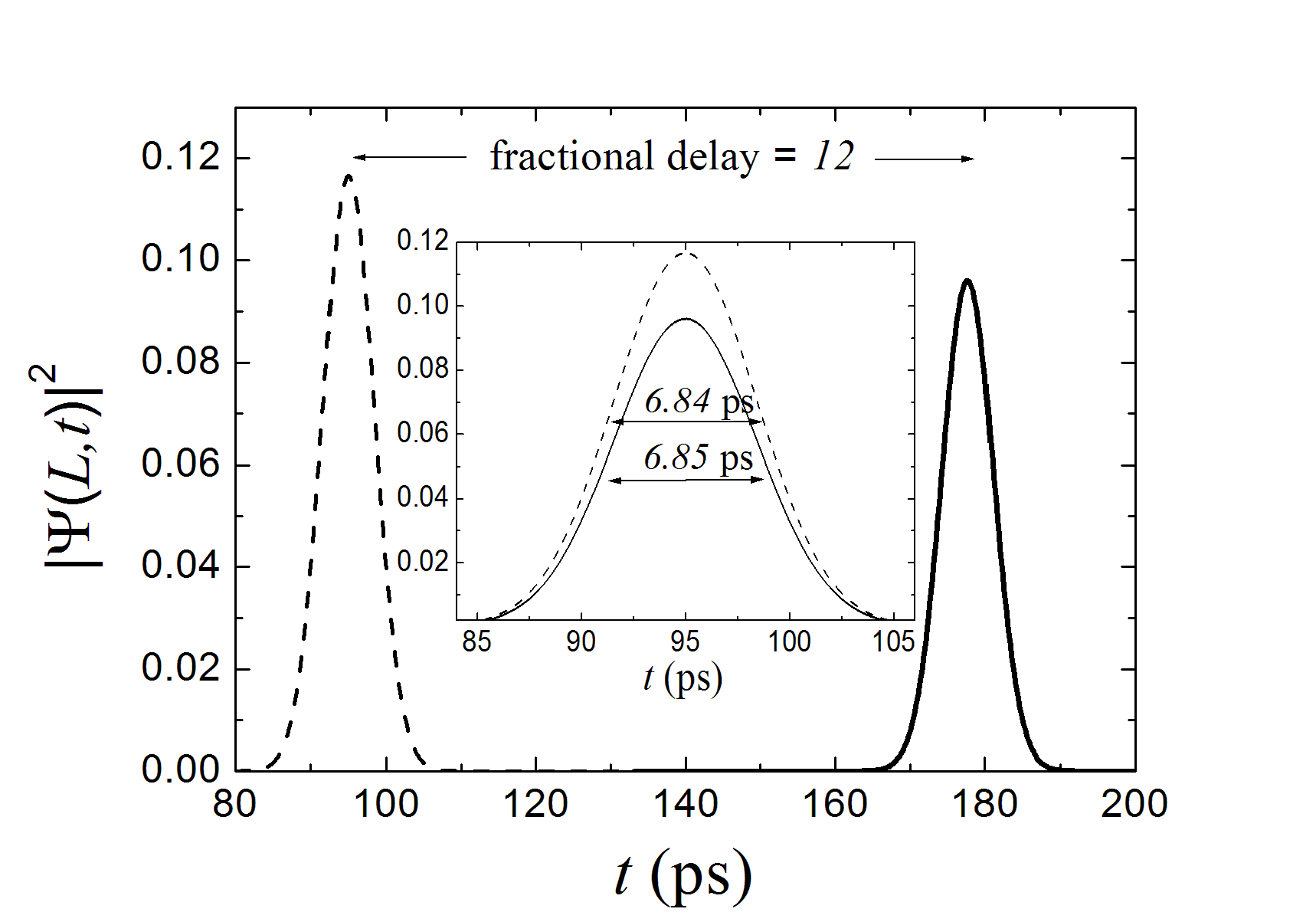}
\caption{Free propagation pulse of width 6.84 ps (dashed line) and
its corresponding transmitted pulse (solid line) for a central
frequency of $\omega_{01} = 1754$ THz which traverses the 1D
grating described in Fig.\ \ref{fig1}.}\label{fig6}
\end{figure}

In view of the previous results, for the high contrasts
considered, it is convenient to use pulses with its central
frequency near the center of the transmission window (so as to
reduce as much as possible the ratio between consecutive peaks),
despite the fact that the group velocity $v_{\rm g}$ is higher in
this region. In Fig.\ \ref{fig6} we compare the shape of a free
pulse of width 6.84 ps (dashed line) with its corresponding
transmitted pulse (solid line). The central frequency of the
incident pulse is $\omega_{01} = 1754$ THz. This is an ultrashort
temporal pulse (bandwidth of 413 GHz) desirable for
telecommunication systems \cite{MORK05,SED07}. We can observe a
large delay of 12 pulses without significant pulse broadening and
high transmission. We will return to this point in the next
section.

\section{Bandwidth dependence of the transmitted pulse parameters}

Let us now study numerically how significant pulse parameters,
such as its transmission coefficient and its fractional delay,
depend on the frequency bandwidth of the incident pulse $2 \Delta
\omega$.

\begin{figure}
\includegraphics[width=.75\textwidth]{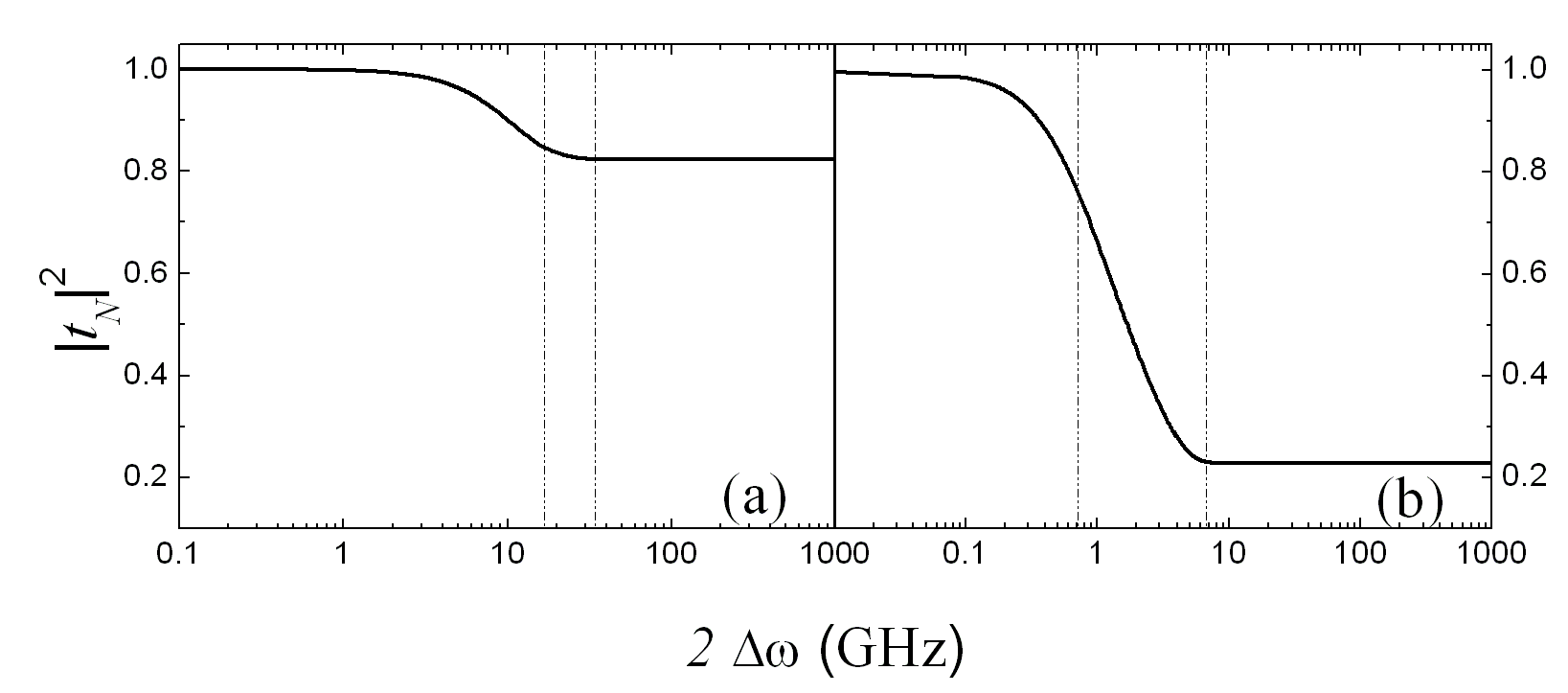}
\caption{Transmission coefficient $|t_N|^2$ as a function of the
bandwidth $2 \Delta \omega$ for an incident pulse whose central
frequency is $\omega_{01} = 1754$ THz (a) and $\omega_{02} = 2225$
THz (b). In both figures, the left dashed line corresponds to the
frequency width of each Lorentzian resonance and the right dashed
line is the lower value of $2 \Delta \omega$ for which the
transmission coefficient $|t_N|^2$ is independent of the frequency
bandwidth.}\label{fig7}
\end{figure}

In Fig.\ \ref{fig7}(a) we represent the transmission coefficient
$|t_N|^2$ as a function of $2 \Delta \omega$ for an incident pulse
whose central frequency is $\omega_{01} = 1754$ THz. The left
dashed line corresponds to the frequency width of the Lorentzian
resonance which is equal to 16.8 GHz (see Fig.\ \ref{fig3}(a))
while the right dashed line (approximately 34 GHz) is the lower
value of $2 \Delta \omega$ for which the transmission coefficient
$|t_N|^2$ is independent of the frequency bandwidth. One observes
that for short bandwidths the transmission coefficient is
practically unity because the frequency pulse is narrower enough
to lay within a transmission resonance. As we increase the
bandwidth $2 \Delta \omega$ the frequency pulse extends over
several resonances what results in a lower transmission
coefficient. When the number of resonances within the frequency
pulse is large enough, the transmission coefficient $|t_N|^2$
reaches a constant value independent of the frequency bandwidth,
in this case 0.86. This is the regime where the traversal time is
well approximated by the classical expression, Eq.\
(\ref{fracdelayclass}). For values of $2 \Delta \omega$ greater
than 34 GHz the transmitted pulse consists of a serie of
decreasing pulses as described in the previous section (see Fig.\
\ref{fig5}(a)) with a short ratio $r = 0.25 \%$.

Similar comments can be applied to Fig.\ \ref{fig7}(b) where the
central frequency of the incident pulse is now equal to
$\omega_{02} = 2225$ THz. The frequency width of the Lorentzian
resonance (left dashed line) is 0.7 GHz (see Fig.\ \ref{fig3}(b))
while the bandwidth limit (right dashed line) is approximately 6.6
GHz. One observes that the transmission coefficient $|t_N|^2$
reaches a lower value than in the previous case. The reason is
that, as explained in Sec.\ II, the overlap between Lorentzians is
negligible near the edge and the transmission in the midpoint
between peaks reduces considerably.

\begin{figure}
\includegraphics[width=.6\textwidth]{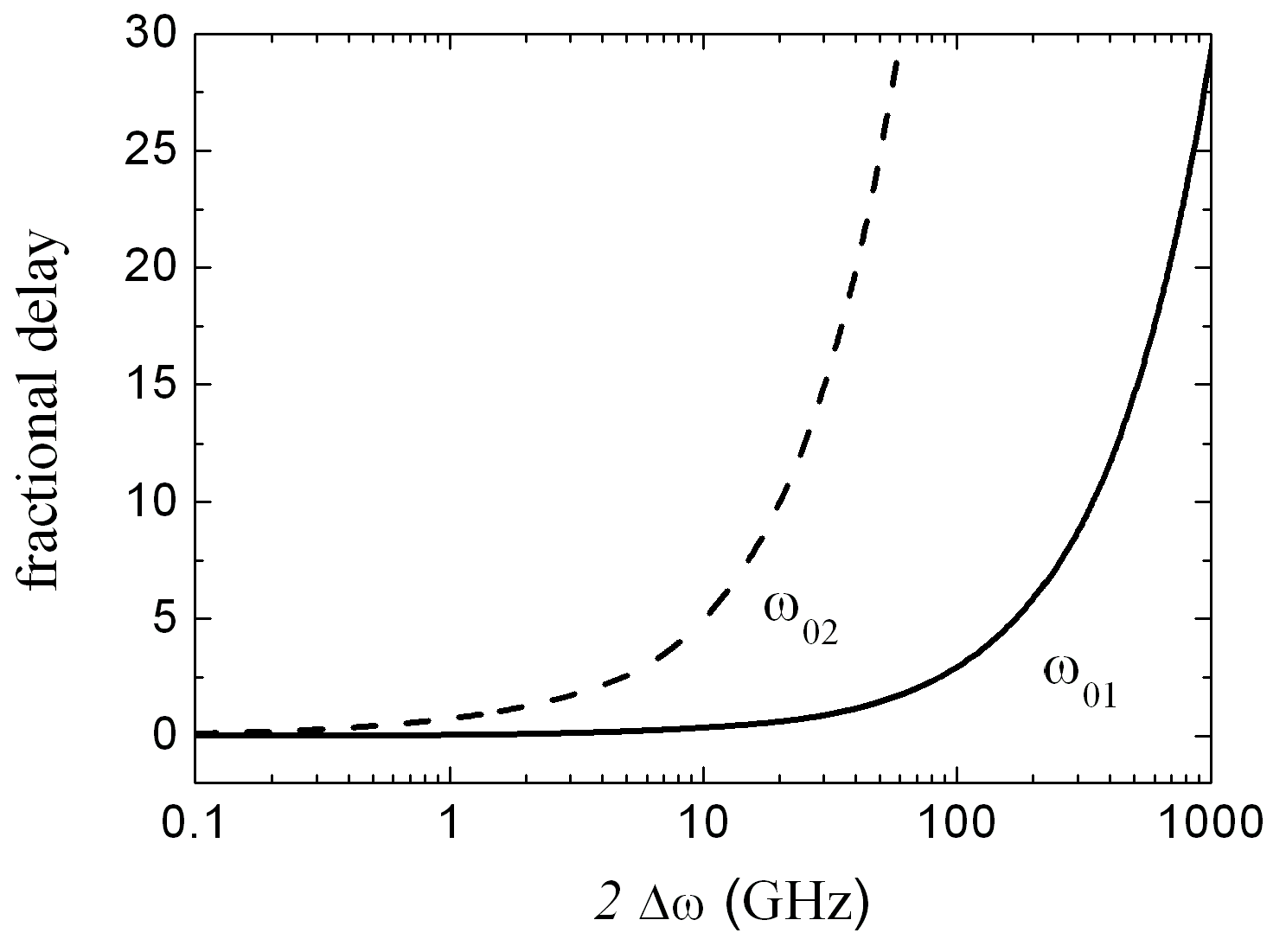}
\caption{Fractional delay versus the frequency bandwidth of the
incident pulse $2 \Delta \omega$ for two different values of the
central frequency, $\omega_{01}$ (continuous line) and
$\omega_{02}$ (dashed line). In the latter case, the ratio $r$ is
quite large.}\label{fig8}
\end{figure}

As regard the fractional delay, we show in Fig.\ \ref{fig8} this
parameter versus $2 \Delta \omega$ for $\omega_{01}$ (continuous
line) and $\omega_{02}$ (dashed line). We have performed these
calculations via Eq.\ (\ref{fracdelaytempus}) though similar
results can be obtained with a classical treatment, Eq.\
(\ref{fracdelayclass}). It can be appreciated that the fractional
delay is, as expected, greater for values of $\omega_{0}$ near the
frequency window edge (where the group velocity is lower). However
the distortion of the transmitted pulse, which is related to the
factor $r$, is considerably high in this region (see Fig.\
\ref{fig5}(b)).

In view of the previous results one can conclude that the
distortion of the transmitted pulse is unacceptable large when its
central frequency lays near the edge of the transmission window.
So, it is convenient to work near the center of this frequency
window, even though it corresponds to larger group velocities.

\section{Discussion and Conclusions}

We have analyzed the transmission properties of pulses through
one-dimensional periodic structures with alternating index of
refraction $n_1$ and $n_2$. We also explored the best conditions
to achieve the maximum delay with the minimum possible pulse
distortion.

As we have already mentioned, in the present work we have not
included either absorption or layer variability. We think that
this is well justified for most situations of interest. In a
typical device employed in slow light experiments, the losses due
to absorption are negligible up to distances of the order of 2 m
\cite{MOK06}. As the structures considered here are 3cm long, we
do not expect any significant effect due to absorption. As regard
of layer variability, it is more difficult to estimate its
effects, but the following argument convinced us that they again
do not need to be considered in first instance. MacGurn \emph{et
al.} \cite{McCh93} studied the localization length associated with
layer variability and concluded that its value decreases (and the
importance of its effects increases) as we approach the band edge.
They performed a numerical simulation for a layer width
variability of 20 \%\ and obtained a localization length of
roughly 20 times the layer period. Lhomm\'e \emph{et al.}
\cite{LC05} estimated that the typical layer width variability in
the type of devices considered by us to be about $0.2$ \%. As in
one-dimensional systems the localization length depends on the
inverse of the square of the disorder, we estimate that the
minimum localization length will be approximately $2\cdot 10^5$
layer spacings. The devices considered have 40000 periods and so
we expect a negligible decrease of the pulse amplitude due to
localization by disorder. Near the band edge localization effects
are potentially more important, due to shorter localization
lengths, so one should be careful with large layer variability in
this region.

For our ideal model, the transmission coefficient $t_N$ can be well approximated by a
sum of Lorentzian resonances and the ratio between their width
$\Gamma_{\rm r}$ and their separation $\Delta\omega_{\rm r}$ is a
significant parameter to characterize the distortion of the
transmitted pulse, which consists of an exponentially decreasing
series of similar pulses given by Eq.\ (\ref{psitr6}). This pulse
distortion is greater near the edge of the transmission window, so
it is convenient to work with pulse frequencies near the center of
this window. Our numerical calculations show fractional delays of
12 pulse widths, without significant distortion and high
transmission, for 6.84 ps width gaussian pulses.

\acknowledgements

The authors would like to acknowledge Vladimir
Gasparian for many interesting discussions. M.O. would like to
acknowledge financial support from the Spanish DGI, project
FIS2006-11126.

\appendix
\begin{appendix}
\renewcommand{\thesection}{A}
\setcounter{subsection}{0}

\section{Transmitted pulse through a periodic arrangement of Lorentzian resonances}

Let us consider that the transmission amplitude of our structure
$t_N(\omega)$ is a periodic function in $\omega$ as given by Eq.\
(\ref{transamplappend}). The transmitted wavefunction $\Psi_{\rm
tr}(t)$ will then be given by
\begin{equation}\label{psitr1}
\Psi_{\rm tr}(t) = C \ \int_{-\infty}^{+\infty} d\omega \exp[-i
\omega t] \ \Phi_{\rm inc}(\omega) \sum_n f(\omega -
\Delta\omega_{\rm r} n, \Gamma_{\rm r}),
\end{equation}
where $C$ is a normalization constant. After introducing the
parameter $v \equiv \omega - \Delta\omega_{\rm r} n$ and
performing some trivial calculations one can write Eq.\
(\ref{psitr1}) as
\begin{equation}\label{psitr3}
\Psi_{\rm tr}(t) = C \ \sum_n \exp[-i \Delta\omega_{\rm r} n t]
\int_{-\infty}^{+\infty} dv f(v, \Gamma_{\rm r}) \Phi_{\rm inc}(v
+ \Delta\omega_{\rm r} n) \exp[-i v t].
\end{equation}

For a large pulse over the frequency domain, the wavefunction
$\Phi_{\rm inc}(\omega)$ is practically constant (so we can
extract this term out of the integral in the previous equation). We
can also express the first exponential in Eq.\ (\ref{psitr3}) as
an integral of the delta function and obtain
\begin{eqnarray}\label{psitr4}
\Psi_{\rm tr}(t) &=& C \ \sum_n \left[\left(
\int_{-\infty}^{+\infty} d\omega \ \delta(\omega -
\Delta\omega_{\rm r} n) \exp[-i\omega t] \ \Phi_{\rm inc} (\omega)
\right) \right.
                \nonumber\\
& \times & \left. \left( \int_{-\infty}^{+\infty} dv f(v,
\Gamma_{\rm r}) \exp[-i v t] \right) \right].
\end{eqnarray}
Taking into account the expression for the Fourier series of a
Dirac comb
\begin{equation}\label{sumdeltas}
\sum_n \delta(\omega - \Delta\omega_{\rm r} n) =
\frac{1}{\Delta\omega_{\rm r}} \sum_n \exp\left[2\pi i \omega
\frac{n}{\Delta \omega_{\rm r}}\right],
\end{equation}
and performing the inverse Fourier transform of the Lorentzian
$f(v, \Gamma_{\rm r})$
\begin{equation}\label{translorent}
\int_{-\infty}^{+\infty} dv f(v, \Gamma_{\rm r}) \exp[-i v t] =
\sqrt{2 \pi} \ \exp\left[-\frac{\Gamma_{\rm r} |t|}{2
\hbar}\right],
\end{equation}
we have for the transmitted wavefunction
\begin{equation}\label{psitr5}
\Psi_{\rm tr}(t) = \frac{\sqrt{2 \pi} C}{\Delta\omega_{\rm r}} \
\exp\left[-\frac{\Gamma_{\rm r} |t|}{2 \hbar}\right] \ \sum_n
\left[\int_{-\infty}^{+\infty} d\omega \exp\left[i \omega
\left(\frac{2 \pi n}{\Delta\omega_{\rm r}} - t \right) \right]
\Phi_{\rm inc} (\omega) \right].
\end{equation}
Identifying the term within brackets in Eq.\ (\ref{psitr5}) as the
inverse Fourier transform of the incident wavefunction $\Psi_{\rm
inc} (t - 2 \pi n / \Delta \omega_{\rm r})$ multiplied by $2\pi$,
we can finally derive the expression for $\Psi_{\rm tr}(t)$, Eq.\
(\ref{psitr6}).

One observes that the transmitted wavefunction $\Psi_{\rm tr}(t)$
is formed by a series of equally spaced peaks, separated by the
inverse of $\Delta \omega_{\rm r}$, and modulated by an
exponentially decaying factor $\exp[-\Gamma_{\rm r} |t|/2 \hbar]$.
The ratio between the intensity of two consecutive peaks $r$ is
then given by Eq.\ (\ref{psitr7}).

\end{appendix}

\bigbreak


\begin{thebibliography}{90}

\bibitem{HAU99} L. V. Hau, S. F. Harris, Z. Dutton and C. H.
Behroozi, Nature {\bf 397}, 594 (1999).

\bibitem{LIU01} C. Liu, Z. Dutton, C. H.
Behroozi and L. V. Hau, Nature {\bf 409}, 490 (2001).

\bibitem{PHI01} D. F. Philips, A. Fleischhauer, A. Mair, R. L.
Walsworth and M. D. Lukin, Phys. Rev. Lett. {\bf 86}, 783 (2001).

\bibitem{OKAW05} Y. Okawachi \emph{et al.}, Phys. Rev. Lett. {\bf 94}, 153902 (2005).

\bibitem{SONG05} K. Y. Song, M. G. Herr\'{a}ez and L. Th\'{e}venaz, Opt. Express {\bf 13}, 82 (2005).

\bibitem{POV05} M. L. Povinelli, S. G. Johnson and J. D. Joannopoulos, Opt. Express {\bf 13}, 7145 (2005).

\bibitem{NOT01} M. Notomi \emph{et al.}, Phys. Rev. Lett. {\bf 87}, 253902 (2001).

\bibitem{GER05} H. Gersen \emph{et al.}, Phys. Rev. Lett. {\bf 94}, 073903 (2005).

\bibitem{MOR04} D. Mori and T. Baba, Appl. Phys. Lett. {\bf 85}, 1101 (2004).

\bibitem{VLA05} Y. A. Vlasov, M. O'Boyle, H. F.
Hamann and S. J. McNab, Nature {\bf 438}, 65 (2005).

\bibitem{MOK06} J. T. Mok, C. M. de Sterke, I. C. M. Littler and B. J. Eggleton, Nature Phys. {\bf 2}, 775 (2006).

\bibitem{SAR06} S. Sarkar, Y. Guo and H. Wang, Opt. Express {\bf 14}, 2845 (2006).

\bibitem{NOV08} I. Novikova, D. F. Phillips and R. L. Walsworth, Phys. Rev. Lett. {\bf 99}, 173604 (2007).

\bibitem{BOYD05} R. W. Boyd, D. J. Gauthier, A. L. Gaeta and A.
E. Willner, Phys. Rev. A {\bf 71}, 023801 (2005).

\bibitem{BA06} O. del Barco, M. Ortu\~{n}o and V. Gasparian,
Phys. Rev. A {\bf 74}, 032104 (2006).

\bibitem{GO95} V. Gasparian, M. Ortu\~{n}o, J. Ruiz and E. Cuevas,
Phys. Rev. Lett. {\bf 75}, 2312 (1995).

\bibitem{AG91} A. G. Aronov, V. Gasparian and U. Gummich, J. Phys.:
Condens. Matter {\bf 3}, 3023 (1991).

\bibitem{GO95a} V. Gasparian, M. Ortu\~{n}o, J. Ruiz, E. Cuevas and M. Pollak,
Phys. Rev. B {\bf 51}, 6743 (1995).

\bibitem{CU96} E. Cuevas, V. Gasparian, M. Ortu\~{n}o and J. Ruiz,
Z. Phys. B {\bf 100}, 595 (1996).

\bibitem{MORK05} J. M\o rk, R. Kjaer, M. van der Poel and K. Yvind, Opt. Express {\bf 13}, 8136 (2005).

\bibitem{SED07} F. G. Sedgwick \emph{et al.}, Opt. Express {\bf 15}, 747 (2007).

\bibitem{McCh93} A. R. McGurn \emph{et al.}, Phys. Rev. B {\bf 47}, 13120 (1993).

\bibitem{LC05} F. Lhomm\'e \emph{et al.}, Appl. Opt. {\bf 44}, 493 (2005).

\end{thebibliography}
\end{document}